\documentclass[aps
,prl,10pt
  ,twocolumn,
twoside,
showpacs,
 nobalancelastpage,
superscriptaddress
,floatfix
 ]{revtex4}

\usepackage{bm}
\usepackage{amsfonts}
\usepackage{amsmath}
\usepackage{amssymb}
\usepackage[latin2]{inputenc}
\usepackage[T1]{fontenc}
\usepackage{epsfig}
\usepackage{mathrsfs}
\usepackage{verbatim}

\newcommand{\sh}{\text{sh}}
\newcommand{\ch}{\text{ch}}

\renewcommand{\i}{{\bm i}}

\newcommand{\e}{\text{e}}
\newcommand{\tr}{\text{tr}}

\newlength{\dinwidth}
\newlength{\dinmargin}
\DeclareMathAlphabet{\scr}{U}{rsfs}{m}{n}

\begin{document}

\title{Quantum superadditivity in linear optics networks: 
sending bits via multiple access Gaussian channels}

\author{Ł. Czekaj}
\affiliation{Faculty of Applied Physics and Mathematics,
Gda{\'n}sk University of Technology, 80-952 Gda{\'n}sk, Poland}

\affiliation{National Quantum Information Center of Gda\'nsk, 81-824 Sopot, Poland}

\author{J. K. Korbicz} 
\email{jkorbicz@mif.pg.gda.pl}
\affiliation{Institute of Theoretical Physics and Astrophysics,
  University of Gda\'nsk, 80-952 Gda\'nsk, Poland}

\affiliation{National Quantum Information Center of Gda\'nsk, 81-824
  Sopot, Poland}

\affiliation{Faculty of Applied Physics and Mathematics, Gda{\'n}sk
  University of Technology, 80-952 Gda{\'n}sk, Poland}

\author{R. W. Chhajlany}

\affiliation{Faculty of Physics, Adam Mickiewicz University,
  Umultowska 85, 61-614 Pozna\'{n}, Poland}

\affiliation{Faculty of Applied Physics and Mathematics, Gda{\'n}sk
  University of Technology, 80-952 Gda{\'n}sk, Poland}

\affiliation{National Quantum Information Center of Gda\'nsk, 81-824
  Sopot, Poland}

\author{P. Horodecki}
\affiliation{Faculty of Applied Physics and Mathematics, Gda{\'n}sk
  University of Technology, 80-952 Gda{\'n}sk, Poland}

\affiliation{National Quantum Information Center of Gda\'nsk, 81-824
  Sopot, Poland}

\begin{abstract}
  We study classical capacity regions of quantum Gaussian
  multiple access channels (MAC).  In classical variants of such
  channels, whilst some capacity superadditivity-type effects such as
  the so called {\it water filling effect} may be achieved, a
  fundamental classical additivity law can still be identified, {\it
    viz.}  adding resources to one sender is never advantageous to
  other senders in sending their respective information to the
  receiver. Here, we show that quantum resources allows
  violation of this law, by providing two illustrative schemes of
  experimentally feasible Gaussian MACs.
 
\end{abstract}


\maketitle

Quantum communication \cite{Nielsen} allows for some phenomena that
are impossible in classical information theory. The natural extension
of the notion of channel to quantum domain has lead to new class of
effects, which can be labeled quantum superadditivities or quantum
activations.  In essence, quantum entanglement allows, under 
certain conditions, for increase in channel capacities (both quantum and
classical) in scenarios that have no classical counterparts, like
e.g. the Butterfly effect \cite{butterfly}.  The known examples of
quantum superadditivities are the following: (i) superadditivity of
the quantum capacity $\mathcal Q$ \cite{Qcap} in multiparty scenario
with an additional classical resource (two-way communication)
\cite{Duer} which is related to the superactivation of bound entanglement
\cite{Shor-superactivation}(see \cite{bound-activation} for the prototype 
activation effect and its extensions \cite{Masanes}) (ii) extremal  
superadditivity i.e.. superactivation of $\mathcal Q$ of ''$0\otimes
0>0$''-type, for a fundamental one sender-one receiver configuration
\cite{Smith} (iii) superadditivity of classical capacity regions for
multiple-access channels (MAC) \cite{Czekaj};
with one of the senders transmitting at a higher rate
than is classically possible, (iv)superadditivity of private capacity 
\cite{Winter-key}, which in particular can be very extensive \cite{Smolin2}.
Other striking alternatives of possible activations have been also provided 
\cite{Smolin1}.

All the above effects have been discovered in finite dimensional
systems. In this Letter we report on a first purely quantum
superadditivity effect for continuous variable (CV) MAC channels,
transmitting classical information (see e.g. Ref.~\cite{Braunstein}
for an overview of quantum information with CV).

We focus on 2-to-1 channels, i.e. two senders transmitting to one
receiver.  Since we will be working with quantum Gaussian MAC channels
(see e.g. Ref.~\cite{Wolf} and \cite{Winter}), we first briefly discuss
classical Gaussian channels  \cite{Cover}: A channel is
modeled by adding  Gaussian noise with variance $N$ to the input data
$X$ with average power $\langle X^2\rangle\leq P$ (required for any CV
channel to prevent unphysical diverging rates).  The maximal mutual
input-output information then reads \cite{Cover} $\max_{\langle X^{2}
  \rangle \leq P} I(X:Y)=C(P/N)$ ($C(x)\equiv 1/2\log(1+x)$).  In the
case of a classical 2-to-1 channel $\Phi$ with input variables
$X_{i}$ constrained by $P_{i}$ ($i=1,2$), the capacity region 
i.e. (closure of) the region of achievable transmission 
rates $R_i$, reads \cite{Cover}:
\begin{eqnarray}
  & &R_{1}\leq \max_{\langle X_{1}^{2} \rangle \leq P_1} I(X_1:Y)=C(P_{1}/N)\label{R1}\\
  & &R_{2}\leq \max_{\langle X_{2}^{2} \rangle \leq P_2} I(X_2:Y)=C(P_{2}/N)\label{R2}\\
  & &R_{1}+R_{2}\leq \max_{\langle (X_{1}+X_{2})^{2}\rangle\leq P_{1}+P_{2}} I(X_1X_2:Y)\nonumber\\
  & &=C\big((P_{1}+P_{2})/N\big).\label{R+R}
\end{eqnarray} 
Here $I(X:Y)=H(X)+H(Y)-H(XY)$ is the mutual information ($H$ stands
for Shannon entropy).

Formulas (\ref{R1}-\ref{R+R}) imply the following rule: {\it maximal
  local rates depend only on local power constraints}.
This is a fundamental property of classical CV channels, which is
weaker than the additivity of discrete ones \cite{Czekaj}, since here
the capacity regions may be nonadditive due to the so called {\it water
  filling effect} \cite{Cover} for inputs with a common power
constraint.  Still, the above rule clearly defines what is classically
forbidden (in both discrete and CV case): {\it adding resources (a
  channel or ,,energy'') to one sender never helps another sender to
  beat his maximal achievable rate.}
 
However, for quantum channels the above rule no longer holds.
In the framework of classical information transmission over a quantum
channel \cite{Ccap,Holevo}, we show that adding a resource---a 1-to-1
channel, to one sender can ``nonlocally'' help beat the other senders
maximal classical transmission rate.  The effect is a CV counterpart
of that of Ref.~\cite{Czekaj}.  It must be stressed that here, we do
not disprove the bound on the {\it total} classical capacity
Eq.~(\ref{R+R}), which is related to the most difficult
formulation of the additivity problem for classical capacities of
quantum channels (see e.g. Ref.~\cite{Shor} and remarkable recent progress
reported in Ref.~\cite{Hastings}).

\begin{figure}
\begin{center}
 \includegraphics[width=0.30\textwidth]{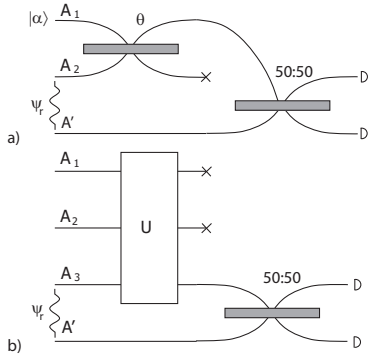}
\caption{Gaussian channels allowing violation 
  of classical bounds on  individual transmission
  rates: a) beam splitter channel; b) triple QND sum gate channel.} \label{setup}
\end{center}
\end{figure}

In both presented schemes, we consider a pair of channels  used
in parallel to communicate to a common receiver: a 2-to-1 MAC channel
$\Phi$ and a 1-to-1 channel $\Psi $ which we shall choose to be the
identity.

In the first example (Fig.~\ref{setup}a), the basic channel
$\Phi:A_1A_2\to B$ is a phase--free beam splitter with transmitivity
$\tau=\cos^2\theta$ and one of its output modes inaccessible to the
receiver:
$\Phi(\varrho_1\otimes\varrho_2)
=\tr_1\left(U_{BS}\varrho_1\otimes\varrho_2U_{BS}^\dagger\right),
$
where $U_{BS}=\exp[i\theta(a_1^\dagger a_2+a_1a_2^\dagger)]$.  The
classical messages $\alpha$, which we assume to be complex numbers,
are encoded into 
coherent states $|\alpha\rangle$. 
The first (second) sender accesses the MAC channel via line
$A_{1}(A_{2})$ while the second sender has lone access to the 1-to-1
channel.

Consider the following protocol: the upper sender sends a strong
coherent state $|\alpha\rangle$ (cf. \cite{Braunstein,Yamamoto}) 
on line $A_1$, while the lower sender
only sends a two-mode squeezed state
\begin{equation}\label{Psir}
|\psi_r\rangle=\exp\left[r(a_2^\dagger a'^\dagger-a_2a')\right]|00\rangle
\end{equation}
on the lines $A_2 A'$,  containing no information. 
  The message $\alpha$ is then decoded through  homodyne detection
  (see Fig. \ref{setup}a).  As shown in Ref.~\cite{Paris}, as
  $\theta\to 0$, $|\alpha|\to \infty$, $|\alpha|\sin\theta=const$, the
  action of $\Phi$ effects a  displacement on the line $A_2$:
  $\Phi(|\alpha\rangle\langle\alpha|\otimes\varrho_2) \to
  D(\alpha\sin\theta)\varrho_2D(\alpha\sin\theta)^\dagger$.  This
  displacement then modulates half of the entangled state present at
  the lower lines. This, combined with the homodyne decoding,
  asymptotically reproduces the CV dense coding protocol
  \cite{CVdense}, which is the crux of our method.

Following the standard approach \cite{Braunstein, Yamamoto}, we assume
Gaussian probability distribution for $\alpha$'s:
\begin{equation}\label{p_0}
p_0(\alpha)=\frac{1}{\pi\sigma^2}\e^{-\frac{|\alpha|^2}{\sigma^2}}.
\end{equation}
Since one ultimately maximizes Eqs.~(\ref{R1}-\ref{R+R}) over $p_0$,
the encoding and the decoding, to show superadditivity it is enough to
consider only a particular scheme, breaking (properly regularized)
bound (\ref{R1}) \cite{bound}.  Using Gaussian character of the
channel and the input, we calculate the mutual information of $A_1$
and $BB'$:
\begin{equation}
I(A_1:BB'|A_2A'=\psi_r)=\log\bigg[1+\frac{\sigma^2\sin^2\theta}{(\ch r-\cos\theta\sh r)^2}\bigg].
\label{I}
\end{equation}
As expected, in the limit $\theta\to 0$, $\sigma\to \infty$,
$\sigma\sin\theta=const$ it reproduces the mutual information of the
dense coding protocol \cite{CVdense}:
$I\to\log(1+\e^{2r}\sigma^2\sin^2\theta)$, with the dispersion
$\sigma^2$ multiplied by the reflectivity of the beam splitter
$\sin^2\theta$ \cite{sin}.

We maximize Eq.~(\ref{I}) under the following power constraints
$\mathcal P$:
\begin{eqnarray}
& & \left\{\text{upper sender av. \# photons}\right\}=\sigma^2\leq N_1,\label{N1}\\
& & \left\{\text{lower sender av. \# photons}\right\}=2\sh^2 r\leq 2N_2.\label{N2}
\end{eqnarray} 
Since all the {\it unconstrained} maxima of $I$ of Eq.~(\ref{I}) are located at a curve:
\begin{equation}\label{maxpoint}
\cos\theta=\text{th}\, r,
\end{equation}
taking into account constraint (\ref{N2}) leads to two cases:
$\cos^2\theta\leq N_2/(N_2+1)$ 
and $\cos^2\theta>N_2/(N_2+1)$. We will study only the first case
here.  Then the constrained maximum of $I$ is attained at
(\ref{maxpoint}) and $\sigma^2=N_1$, and reads \cite{case2}:
\begin{equation}\label{Imax}
I_{max}^{ent}=\log\left(1+N_1\right).
\end{equation}
Note that, due to Eq.~(\ref{maxpoint}), $I_{max}^{ent}$ does not depend
on $N_2$ or $\theta$ explicitly.
In fact, Eq.~(\ref{maxpoint}) connects the signal parameter (squeezing
$r$) with the device parameter (transmitivity $\cos^2\theta$).
Furthermore, one might expect $I_{max}^{ent}$ to scale like the dense
coding capacity, $I\sim\log N^2$, while from Eq.(\ref{Imax}) it scales
only like $\log N$, i.e. just as for the standard Fock-state, coherent
state, and squeezed-state encodings \cite{Yamamoto}. 
This apparently linear regime is due to the fact that the channel
is highly lossy (most of $A_1$'s energy 
is lost;  see Fig.\ref{setup}a), while the quoted
schemes assume loss-less transmission.

To show superadditivity, it is enough to prove that under the same
constraints (\ref{N1}-\ref{N2}), the single-shot quantity
$I_{max}^{ent}$ of Eq.~(\ref{Imax}) surpasses the regularized
\cite{regularization} quantity $\max_{\mathcal P}I(A_1:B|A_2)$, which
we call $\mathcal I_{max}^{prod}$.  This stems from the fact that
regularization can only increase the capacity region, due to
entanglement between the copies.  We first employ a chain of rather
rough bounds on the related mutual information:
$I(A_1:B|A_2)\leq I(A_1A_2:B)\leq C(\Phi)$, where $C(\Phi)$ is the
single-shot classical capacity of $\Phi$. By the famous Holevo theorem
\cite{Holevo}, $C(\Phi)$ is in turn bounded by the maximum output
entropy 
($S(\varrho)$ stands for the von Neumann entropy of $\varrho$):
\begin{eqnarray}
& &C(\Phi)\leq \text{max}_{\varrho}
S\big(\Phi(\varrho)\big)=g(N_{out})\nonumber\\
& &\equiv-N_{out}\log N_{out}+ (1+N_{out})\log(1+N_{out})\label{Smax}
\end{eqnarray}
which for given average output photon number $N_{out}$ is attained for
the corresponding thermal number state, whose entropy is $g(N_{out})$.
The photon number $N_{out}$ can be easily bounded, since we allow only
product inputs subject to (\ref{N1}-\ref{N2}):
\begin{equation} \label{Nout}
N_{out}\le N_{max}(\theta)\equiv\big(\sqrt{N_1}\sin\theta+\sqrt{2N_2}\cos\theta\big)^2
\end{equation}
(the lower sender can pump {\it all} of his available energy, $2N_2$,
into $A_2$).

Next, we consider $n$-copies of $\Phi$ and show that the bound
(\ref{Smax}-\ref{Nout}) is additive.  Power constraints
(\ref{N1}-\ref{N2}) in this case read, $\mathcal P ^{\otimes n}:
\sum_{k=1}^n N_1^{(k)}=nN_1$, $\sum_{k=1}^n N_2^{(k)}=nN_2$, where
$N_i^{(k)}$ is the average photon number at the $i$-th input of the
$k$-th copy.  Using the same result of Ref.~\cite{Smin} as above, we
obtain $\max_{\psi} S(\Phi^{\otimes
  n})=S(\varrho_G(N_{out}(\Phi^{\otimes n})) \leq \sum_k
S(\varrho_G^{(k)}( N_{out}^{(k)}))$; $\varrho_G$ is some $n$-mode
Gaussian state with a total of $N_{out}(\Phi^{\otimes n})$ photons,
$\varrho^{(k)}_G$ is its $k$-th copy reduction with $N^{(k)}_{out}$
photons.  In the last step we used subadditivity of the von Neumann
entropy.
Using Eq.~(\ref{Nout}) for each output, we obtain $N_{out}^{(k)}\leq
N_{max}^{(k)}(\theta)=
\left(\sqrt{N_1^{(k)}}\sin\theta+\sqrt{2N_2^{(k)}}\cos\theta\right)^2$.
We further take the maximum over all possible partitions of the input
energy.  Maximum of $\sum_k N_{max}^{(k)}(\theta)$ under $\mathcal
P^{\otimes n}$ is easily calculated, noting that for a length of a
sum of two vectors is maximal when they are parallel.  This
corresponds to the equal partition of the input energy:
$N_1^{(k)}=N_1$, $N_2^{(k)}=N_2$ for all $k$, so that $\sum_k
N_{max}^{(k)}(\theta) =nN_{max}(\theta)$, which finally leads to
$(1/n)\max_{\psi} S(\Phi^{\otimes n})\leq g\big(N_{max}(\theta)\big)$.

In order to identify superadditivity, we first analyze the
characteristics of the considered channel in the regime where quantum
effects are most prominent, {\it i.e.} when the operation of the setup
is closest to the dense coding protocol. Assuming thus the largest
transmitivity allowed 
 (see also \cite{max})
\begin{equation}\label{regime}
\cos^2\theta=\frac{N_2}{N_2+1} 
\end{equation}
leads to  the desired upper bound (Eq.(\ref{Smax})):
\begin{equation}
\mathcal I_{max}^{prod}\leq g({\bm N}_{max}),
\label{bound}
\end{equation}
with ${\bm N}_{max}=\left(\sqrt{N_1}+\sqrt{2}N_2\right)^2/(N_2+1)$.
In what follows, for comparison with the entanglement assisted
capacity (\ref{Imax}), we consider the worst case scenario for our
scheme {\it i.e.}, equality in Eq.(\ref{bound}). The "capacity
enhancement parameter" $I_{max}^{ent}/\mathcal I_{max}^{prod}$ 
attains its maximum value under the condition
$\sqrt{N_{1}} = \sqrt{2}(N_{2}+2)$ (leading to ${\bm N}_{max}=
8(N_{2}+1)$. The maximum value of this parameter is therefore given by
(see Eqs.(\ref{Imax}) and (\ref{Smax})) $\log
\Big(1+2(N_{2}+2)^2\Big)/g(8(N_{2}+1))$. 
In particular, the enhancement over product state capacity approaches
2 as $N_{2} \rightarrow \infty $,  
clearly proving quantum superadditivity in this scheme.

\begin{figure}
\begin{center}
  \includegraphics[width=0.5\textwidth]{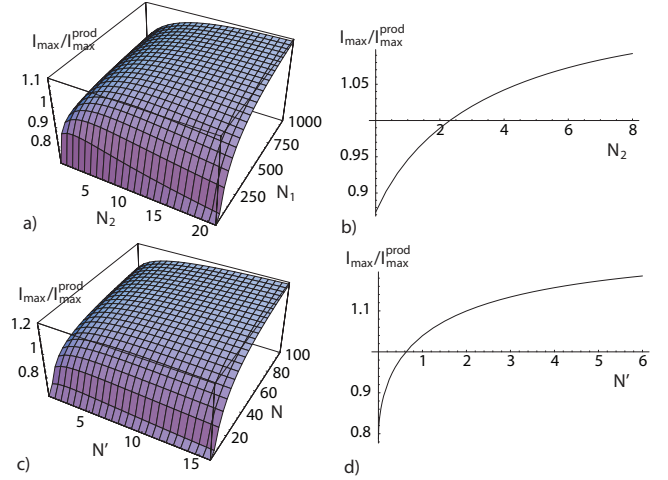}
\caption{Ratios of the classical capacities of
entanglement assisted channels and regularized capacities of the same 
channels with product inputs as functions of the power constraints 
for a) beam-splitter channel of Fig.~\ref{setup}a; b) the cut for $N_1=1000$;
and c) triple QND sum gate channel of Fig.\ref{setup}b; d) the cut for $N=100$. 
}\label{I/I}
\end{center}
\end{figure}

 More generally, superadditivity occurs for appropriately large powers
 of the inputs of both senders (Fig.\ref{I/I}a).  For example,
 Fig.~\ref{I/I}b shows the capacity enhancement for the average
 coherent input power $N_1=1000$. Note that each point in the figure is
 in fact a distinct physical channel---different $N_2$'s correspond via
 Eq.~(\ref{regime}) to different beam splitters.  Superadditivity
 appears for $N_2\ge 2.3105$ which corresponds to a squeezing $r \geq
 10.47$ dB of the lower sender's signal state.

 Our second example (Fig.~\ref{setup}b) consists of the Gaussian MAC
 channel $\Phi:A_1A_2A_3\to B$ defined by
 $\Phi(\varrho_{12}\otimes\varrho_3)=
 \tr_{12}\left(U\varrho_{12}\otimes\varrho_3U^\dagger\right)$, where
\begin{eqnarray}
U=\exp\left[-i( \hat{x}_1 \hat{p}_3-\hat{p}_2\hat{x}_3)\right]. 
\label{unittransf}
\end{eqnarray}
This unitary is generated by two quantum nondemolition (QND) type
interactions and can be decomposed into three QND sum gates
\cite{qndsum} between modes $1,2,3$. The upper sender holds both lines
$A_1$ and $A_2$, while the lower one holds lines $A_3$ and $A'$ of
channel $\Psi$.

The transmission protocol is the following. The upper sender encodes a message
$\alpha=\alpha_R+\i\alpha_I$ into the displaced  state:
$|\psi_{in}\rangle_{A_1A_2}=D(\alpha_R,0)|R\rangle\otimes
D(0,\alpha_I)|-R\rangle, $
where $|\pm R\rangle$ are one mode squeezed vacuum states with
squeezing parameters $\pm R$.  For large squeezing $R\to\infty$,
the action of $\Phi$ again approaches the displacement $D(\alpha)$.
Just as before, we assume that: i) the lower sender always sends a two
mode squeezed state $\psi_r$, cf. Eq.~(\ref{Psir}), on the lines
$A_3A'$; ii) the input probability $p_0(\alpha)$ is given by
Eq.~(\ref{p_0}); iii) the decoding is done through  homodyne
detection on the output lines $BB'$ (see Fig.~\ref{setup}).

The mutual information between the upper sender $A_1A_2$ and the
receiver $BB'$ then reads:
\begin{equation}
  I(A_1A_2:BB'|A_3A'=\psi_r)=\log\left[1+
    \frac{\sigma^2}{e^{-2r}+(e^{-2R}/2)}\right] \label{PhiChannelCap}
\end{equation} 
As expected, in the limit $R\to\infty$ we again recover the CV dense
coding capacity \cite{CVdense}.  We apply similar photon number
constraints $\mathcal P$ as before in Eqs.~(\ref{N1}-\ref{N2}):
\begin{eqnarray}
&\left\{\text{upper sender av. \# photons}\right\}&\nonumber\\
&=\sigma^2 + 2\sh^2 R \leq N,&\label{PhiCnsrt1}\\
&\left\{\text{lower sender av. \# photons}\right\}\nonumber &\\
&=2\sh^2r\leq 2N'.&\label{PhiCnsrt2}
\end{eqnarray}
The constrained maximum of $I$ of Eq.~(\ref{PhiChannelCap}) is
achieved when the inequalities (\ref{PhiCnsrt1}-\ref{PhiCnsrt2}) are
saturated: $\sh^2r=N'$ and $\sigma^2=N-2\sh^2R$, the latter leading
through maximization of Eq.~(\ref{PhiChannelCap}) to $2\e^{2R}=-e^{2r}
+ \sqrt{e^{4r} + 4e^{2r}(N+1)+4}$.  Substituting the above values back
into Eq.~(\ref{PhiChannelCap}) yields the desired maximum achievable
one-shot rate $I_{max}^{ent}$.
 
Using the same argument as in the previous scheme, we compare
$I_{max}^{ent}$ with the classical capacity of $\Phi$, which 
is bounded by Eq.~(\ref{bound}) now with
\begin{equation}
{\bm N}_{max}=\left(\sqrt{2N' + \frac{1}{2}}+\sqrt{N+1}\right)^2 - \frac{1}{2}. 
\end{equation}
The above bound follows directly from the input-output relations for
the MAC channel (\ref{unittransf}), constraints
(\ref{PhiCnsrt1}-\ref{PhiCnsrt2}) for product inputs in the cut
$A_1A_2|A_3$. The regularization procedure follows exactly as in the
beam splitter channel case, yielding additivity of the above bound.
Regions of superadditivity are manifest in the plot of
$I_{max}^{ent}/\mathcal I_{max}^{prod}$ in Fig.~\ref{I/I}c.
Fig.~\ref{I/I}d shows the cut for $N=100$,
cf. Eq.~(\ref{PhiCnsrt1}). Quantum superadditivity occurs for power
$N'\geq 0.63$ corresponding to $6.33$ dB of two-mode squeezing while
the upper sender uses $4.21$ photons in each line requiring
one-mode squeezing of $12.73$ dB. In the end of the range i.e. $N'=6$
(noise reduction $14.15$ dB) the upper sender uses at most $9.45$
photons per line requiring $16.02$ dB.

We have thus shown capacity superadditivity in Gaussian MAC channels,
that has no classical analog. Finally, we comment on perspectives for
implementation of proof-of-principle experiments of these effects. The
first scheme consists of an extremely basic linear optics setup, while
a QND sum gate has also been implemented \cite{yoshikawa}. The main
obstacle for observation of superadditivity effects is the amount of
squeezing required. However, recently techniques yielding squeezing of
up to 10 dB have been reported with 15 dB being claimed to be
attainable in the near future \cite{expsq}. Such parameters would be
sufficient for superadditivity as shown in this paper.

We thank M. \& R. Horodecki for discussions and the EU integrated
project SCALA for financial support. RWC gratefully acknowledges
support from the Foundation for Polish Science (FNP).


\begin{thebibliography}{99}
\bibitem{bound} Since the upper sender has no access to $\Psi$, his rate is 
classically bounded by (\ref{R1}).
\bibitem{sin} This is so since in the considered limit 
$\Phi(|\alpha\rangle\langle\alpha|\otimes\cdot)$ approaches
$D(\alpha\sin\theta)$.
\bibitem{case2} In the second case , the maxima (\ref{maxpoint}) lie outside
the region (\ref{N1}-\ref{N2}), so that the constrained 
maximum of $I$ is simply attained for $\sh^2r=N_2$ and $\sigma^2=N_1$:
$I_{max}^{ent}= \log\big[1+N_1\sin^2\theta/2(\sqrt{N_2+1}-\cos\theta\sqrt{N_2})^2\big]$.
\bibitem{regularization} For a quantity $I(\Phi)$ its regularization 
$\mathcal I$ is defined as $\mathcal I=\lim_{n\to\infty}(1/n) I(\Phi^{\otimes n})$.
\bibitem{max} One might think that the bound (\ref{Nout}) has to be further 
maximized over $\theta$, yielding $N_1+2N_2$, which is
larger that $N_1+1$ (cf. Eq.~(\ref{Imax}). However, we stress that for
our purpose it is enough to find {\it some} $\theta$, for which 
$I_{max}^{ent}\ge\mathcal I_{max}^{prod}$.
\end{thebibliography}
\end{document}